# Single-nanoparticle detection using quasi-bound states in the continuum supported by silicon metasurfaces

Keisuke Watanabe[1]*, Samuel Crowther[2], Masanobu Iwanaga[3], Frank Vollmer[2], and Tadaaki Nagao[1,4]

1.* International Center for Materials Nanoarchitectonics (MANA), National Institute for Materials Science (NIMS), 1-1 Namiki, Tsukuba, Ibaraki 305-0044, Japan.
Email: watanabe.keisuke@nims.go.jp. https://orcid.org/0000-0002-4285-2135
2. Department of Physics and Astronomy, Living Systems Institute, University of Exeter, EX4 4QD, Exeter, UK.
3. Research Center for Electronic and Optical Materials, National Institute for Materials Science (NIMS), 1-1 Namiki, Tsukuba, Ibaraki 305-0044, Japan.
4. Department of Condensed Matter Physics, Graduate School of Science, Hokkaido University, Kita 10, Nishi 8, Kita-ku, Sapporo 060-0810, Japan.

**Abstract:** The detection of single particles or molecules represents a critical milestone in the development of biosensing technologies. Recently developed optical sensors based on quasi-bound states in the continuum (qBICs) have primarily focused on detecting global refractive index changes, aiming to simultaneously enhance both refractive index sensitivity and quality ($Q$) factors. However, sensors capable of resolving local refractive index perturbations, such as the binding of a nanometer-sized molecule on a surface, remain elusive and have not yet been demonstrated in BIC geometries due to the limited $Q$ factors and relatively large mode volumes. Here, we demonstrate low-contrast BIC metasurfaces that can perform sensing with a virus-sized single-nanoparticle resolution. The qBIC resonance operating at the critical coupling condition exhibits an experimental $Q$ factor of $4.5 \times 10^4$ in heavy water. The strong interaction between the localized electric field and polystyrene nanoparticles with a diameter of 100 nm enable the experimental observation of step-like resonance wavelength shifts, serving as signatures of individual particle binding events. Furthermore, binding-induced modifications to the qBIC resonance alter the optical confinement and asymmetry factor, inducing changes not only in the resonance wavelength but also in the linewidth and amplitude with single-particle sensitivity. Combined with position-insensitive response and free-space accessible features, low-contrast BIC metasurfaces provide a user-friendly platform for next-generation single-molecule sensing integrated with microfluidic systems.

Quasi-bound states in the continuum (qBICs) have gained considerable research interest across a wide range of applications because of their high quality ($Q$) factors and strong electric field enhancement.[1] The resulting enhancement of light–matter interactions has emerged as a central topic of interest[2–6], particularly for sensing applications in aqueous environments. Thus far, several label-free qBIC-based sensor geometries have been investigated, primarily by exploiting resonance wavelength shifts induced by changes in the refractive index of the surrounding medium.[7–11] The symmetry breaking of symmetry-protected BICs emerging at the Γ point of the Brillouin zone offers a significant advantage, namely, sharp resonances can be accessed under normal-incidence excitation from free-space.

Conventional design strategies for these sensors aim to simultaneously maximize $Q$ factors and refractometric sensitivity for capturing global, large-area refractive index variations in the surrounding medium near the surface. In this framework, the confinement factor of the optical mode into the external medium $\Gamma_{ext}$, which is directly proportional to the global refractive index sensitivity, is a critical parameter. However, increasing $\Gamma_{ext}$ typically degrades the $Q$ factor. Recent experimental approaches have addressed this trade-off using nanogap geometries,[12] hybrid plasmonic–dielectric metasurfaces,[13] and Bloch surface waves.[14] On the other hand, sensors that can resolve local refractive index perturbations, such as the binding of nanometer-sized molecules on a surface,[15] remain elusive and are not yet to be realized in BIC metasurfaces.

The fractional frequency shift induced by a nanometer-sized molecule follows $\Delta\omega/\omega \propto -\alpha_{ex}/V_{eff}$, where $\alpha_{ex}$ and $V_{eff}$ represent the excess polarizability of the molecule and effective mode volume,[16] respectively. Therefore, reducing the mode volume is essential for achieving large resonance shifts. When combined with high-resolution wavelength tracking afforded by high $Q$ factors, optical resonators with high $Q/V$ ratios are promising platforms for sensors with single-molecule resolution. Over the past decade, various single-particle/molecule optical sensors have been developed,[17–21] including whispering gallery mode (WGM) resonators based on microspheres,[22] microtoroids,[23] microrings[24], and microlasers[25] exhibiting ultrahigh $Q$ factors. Further, photonic crystal cavities confining light to diffraction-limited volumes have enabled the detection of single viruses[26] and single molecules.[27] In comparison, conventional BIC metasurfaces exhibit lower $Q$ factors because qBIC resonances are susceptible to fabrication imperfections, and the scattering losses arising from the surface roughness restrict the potential to achieve ultrahigh $Q$ factors. This limitation prevents BIC metasurfaces from reaching the signal-to-noise required for single-molecule detection. To overcome this, we recently demonstrated a BIC metasurface with $Q$ factors exceeding 100,000 using a low-contrast geometry.[28] This ultrahigh $Q$ factor arises from the intrinsically high radiative $Q$ factors of higher-order qBICs combined with reduced scattering losses in shallow-etched geometries. These features suggest that low-contrast BIC metasurfaces can provide unprecedentedly high $Q/V$ values, making them highly sensitive to local refractive index perturbations.

In this study, we demonstrate the real-time detection of virus-sized single nanoparticles in water using low-contrast BIC metasurfaces with ultrahigh $Q$ factors. We fabricate shallow-etched silicon metasurfaces with an asymmetry parameter satisfying the critical coupling condition, enabling the observation of discrete, step-like resonance wavelength shifts as signatures of the surface adsorption events of individual polystyrene (PS) particles. We analyze the overlap between localized electric fields and nanoparticles, revealing the largest wavelength shifts for particles with a 100 nm diameter. Furthermore, we observe binding-induced changes in linewidth and resonance amplitude. These variations can be attributed to the highly sensitive nature of qBIC resonances to symmetry perturbation, representing a unique sensing mechanism enabled by our dielectric BIC metasurfaces for single-molecule detection.

■ **Results and discussion**
**Structural and sensing characteristics**
The low-contrast BIC metasurfaces were fabricated on a silicon-on-insulator (SOI) wafer featuring a 400-nm-thick top silicon layer. Shallow pair-rod nanostructure arrays were formed by dry etching only approximately 52 nm of the top silicon layer with a period $P$ of 760 nm. Figure 1a illustrates PS particles dispersed around a low-contrast BIC metasurface. Tunable laser light is focused onto the metasurface, and the transmission spectrum is recorded via a photodiode. When a single nanoparticle lands on the surface, it interacts with the localized electric field, inducing a resonance wavelength shift (Figure 1b). By continuously tracking the resonance peak, adsorption and desorption events can be observed as discrete, step-like changes in the signal. The fabricated metasurface is integrated into a polydimethylsiloxane (PDMS) microfluidic channel (Figure 1c), and the transmission spectrum is monitored while injecting a PS particle dispersion. Figure 1d shows a scanning electron microscope (SEM) image of the fabricated BIC metasurface, wherein each unit cell includes a pair of rods with lengths $L + 2\Delta L$ and $L - 2\Delta L$. The designed rod length $L$ and width are 640 nm and $L - P/2 = 260$ nm, respectively. Introducing this length asymmetry transforms the symmetry-protected BIC into a qBIC with an asymmetry parameter $\alpha = 2\Delta L/L$, enabling sharp qBIC resonances to be experimentally accessed under normal-incidence excitation.

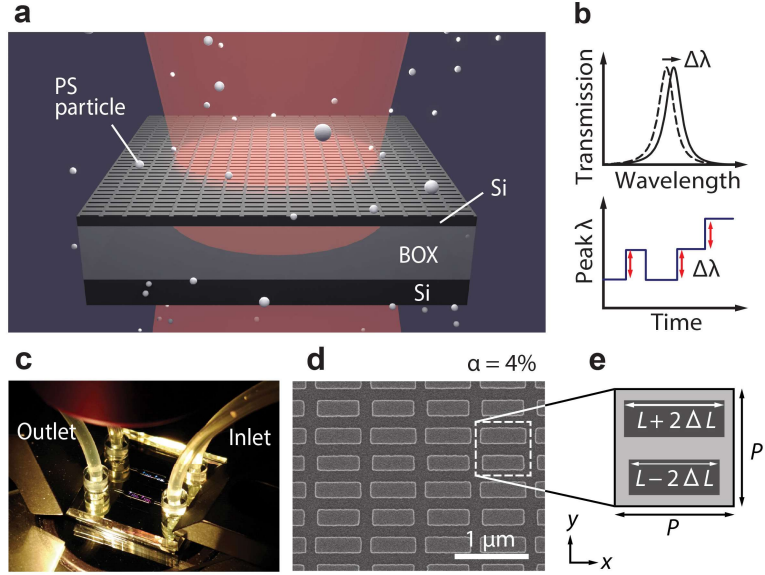

**Figure 1. Low-contrast BIC metasurfaces for single-nanoparticle detection.** (a) Schematic of single PS nanoparticle detection using a low-contrast BIC metasurface fabricated on an SOI wafer. (b) Sensing mechanism. The resonance wavelength shift (top) and its temporal evolution (bottom) displaying step-like signal changes. (c) PDMS microfluidic channel integrated on a metasurface array. (d) SEM image of a fabricated BIC metasurface with an etching depth of approximately 52 nm and asymmetry parameter α = 4%. The dashed region indicates a unit cell. (e) Structural parameters of a single unit cell.

Figure 2a details the custom-built experimental setup. A tunable laser (TSL210V, santec) with a tuning range of 1500 to 1600 nm was focused onto the sample through a 10× objective lens (M Plan Apo NIR, NA = 0.26, Mitutoyo) and a polarizer. The focused laser beam diameter was evaluated to be 28.4 μm using the knife-edge method (Supplementary Information S1). The transmitted beam was detected using an InGaAs picowatt receiver (WPR-2K-IN-FST, FEMTO), with the background noise minimized using a cross-polarized configuration implemented with an additional polarizer placed in front of the photoreceiver. Figure 2b shows the transmission spectrum for a metasurface with α = 4% measured using the electromechanical wavelength scanning of the tunable laser. High-resolution wavelength tuning was achieved using the built-in piezoelectric control of the laser wavelength by applying a 10 Hz triangular voltage generated by a function generator (AFG1022, Tektronix). The resulting resonance spectrum synchronized with the triangular modulation was recorded using a data acquisition (DAQ) device (USB-6212, National Instruments) controlled through a LabVIEW interface. The acquired spectrum (inset of Figure 2b) was fitted with a Fano function expressed as $F(\delta) = A_0 + F_0(q + \delta)^2/(1 + \delta^2)$, where $A_0$ and $F_0$ represent fitting constants and $q$ represents the Fano asymmetry parameter. Here, $\delta = (\omega - \omega_0)/\gamma$, where $\omega_0$ and $\gamma$ represent the resonance frequency and damping rate, respectively. The fitting yielded an experimental $Q$ factor (= $\omega_0/2\gamma$) of approximately $4.5 \times 10^4$. All the measurements were performed in heavy water ($D_2O$) to suppress unwanted overtone absorption from water. The laser wavelength for piezo scanning was calibrated using an optical spectrum analyzer (AQ6374E, Yokogawa), resulting in a slight offset in the displayed wavelengths (horizontal axis) between motor-driven and piezo scanning. Figure 2c shows the $Q$-factor analysis performed by measuring experimental $Q$ factors for the different values of α (Supplementary information S2 for the transmission spectra for all α) and comparing them with theoretical fitting curves. The radiative $Q$ factors of qBIC resonances, $Q_r$, which follow the relationship $Q_r = Q_0 \alpha^{-2}$, were calculated at different α using finite-difference time-domain (FDTD) simulations, yielding $Q_0 = 128.9$. For a small α, the experimental $Q$ factors deviate significantly from $Q_r$. This can be attributed to scattering losses arising from fabrication imperfections. These losses introduce an additional contribution $Q_{scat}^{-1}$, limiting the total $Q$ factor as $Q^{-1} = Q_r^{-1} + Q_{scat}^{-1}$. The experimental $Q$ factors fitted well with this equation using nonlinear least squares curve-fitting method, resulting in $Q_{scat} = 7.8 \times 10^4$. Recently, we demonstrated that optimal balance between a high $Q$ factor and large resonance amplitude can be achieved under a critical coupling condition, where the radiative and nonradiative $Q$ factors are equal, thereby minimizing the effect of wavelength fluctuations.[29] In the present study, the critical coupling condition was experimentally determined to occur at $\alpha_{CC}$ = 4%.

Next, unfunctionalized PS particles ($d$ = 500 nm) dispersed in $D_2O$ with 100 mM NaCl were injected into the PDMS microfluidic channel, and the resonance peak wavelength was monitored in real-time. As shown in Figure 2d, multiple discrete wavelength steps were observed. Following the initial step-like shift, the resonance wavelength exhibits up-and-down oscillations; this behavior is attributed to the particle transiently bouncing near the surface because of weak trapping under the thermal fluctuations. The subsequent negative step suggests the particle moved away from the surface. A distinct step with reduced noise was then observed around 130 s, suggesting the permanent adsorption of the particle onto the surface. The observation of both weak trapping and irreversible binding within a single experiment suggests the presence of electrostatic repulsion between the anionic sulfonate ($R-SO_3^-$) groups on the PS particle[30] and negatively charged native oxide ($SiO_2$) surface in water, even though this repulsive interaction is partially screened by reducing the Debye length through the addition of salt (100 mM NaCl).[31]

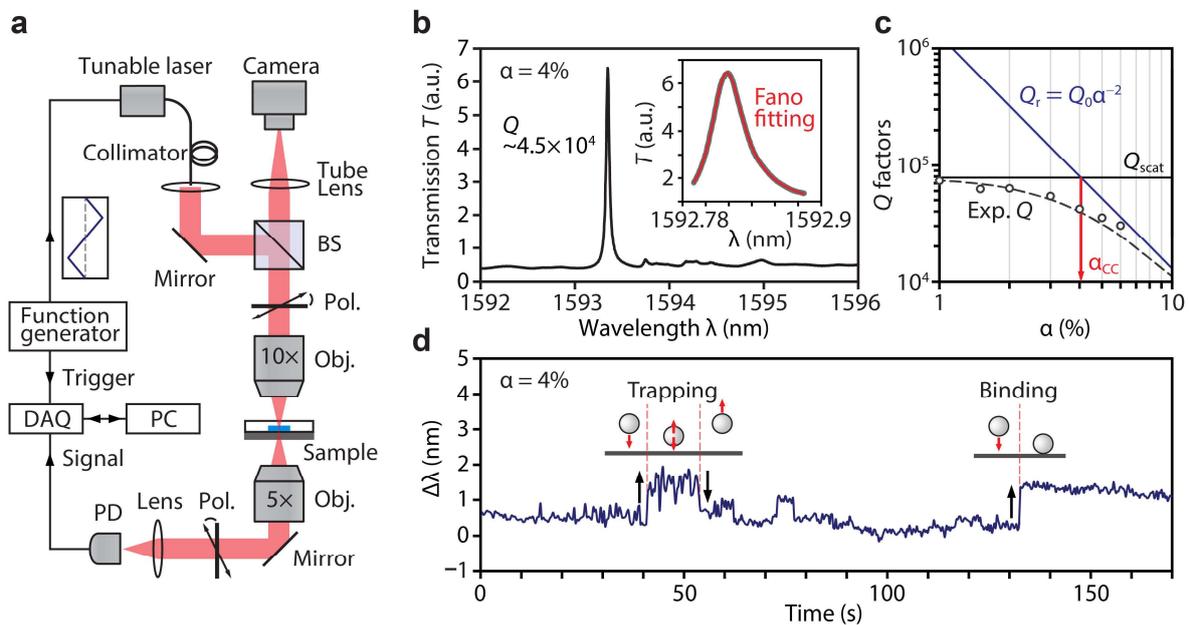

**Figure 2. Optical characterization and single-particle sensing performance.** (a) Schematic of high-resolution transmission spectral measurement setup. (b) Representative transmission spectrum measured in $D_2O$ for a fabricated BIC metasurface with asymmetry parameter $\alpha$ = 4%. Inset shows Fano fitting obtained under fine wavelength tuning based on piezoelectric laser control. (c) $Q$-factor analysis showing experimental $Q$ factors (open circles) fitted as total $Q$ factors (black dashed line) with simulated radiative $Q$ factor (blue solid line) and scattering $Q$ factor (black solid line). (d) Real-time resonance wavelength shift measured in a dispersion of unfunctionalized PS particles ($d$ = 500 nm) showing both weak trapping and discrete step-like binding events.

**Size dependence**
We conducted FDTD simulations of the electric field enhancement under normally incident $x$-polarized plane-wave to clarify sensing characteristics associated with PS particles binding to low-contrast BIC metasurfaces (Figure 3a). As indicated by the electric field distribution in the $yz$-plane, two antinodes appear along the $z$-direction. This corresponds to a higher-order mode ($TE_{122}$) qBIC strongly confined within the top silicon layer and exhibits a high radiative $Q$ factor. In the $xy$-plane at the shallow-etched surface, the electric field is localized at the edges of both rods, which are exposed directly to the external medium (Figure 3b). The localized fields then decay exponentially along the $z$-direction from the silicon surface into the surrounding water (Supporting Information S3 for more details). Such field localization at the interface while preserving a high $Q$ factor presents a key advantage for sensing applications that target molecular binding events. Based on these results, we classify two distinct nanoparticle binding configurations based on particle diameter $d$ and gap spacings between adjacent rods, denoted as $l_1$ (upper rods) and $l_2$ (lower rods), across neighboring unit cells (Figure 3c). When $d < l_1, l_2$, nanoparticles can access the gap region at the shallow-etched surface and cause strong interactions with the localized electric field (configuration (i)). When $d > l_1, l_2$, nanoparticles are excluded from gap regions, which results in weaker field interactions (configuration (iii)). In both cases, resonance wavelength shifts can also arise from particle binding on the top surfaces of the rods (configurations (ii) and (iv)).

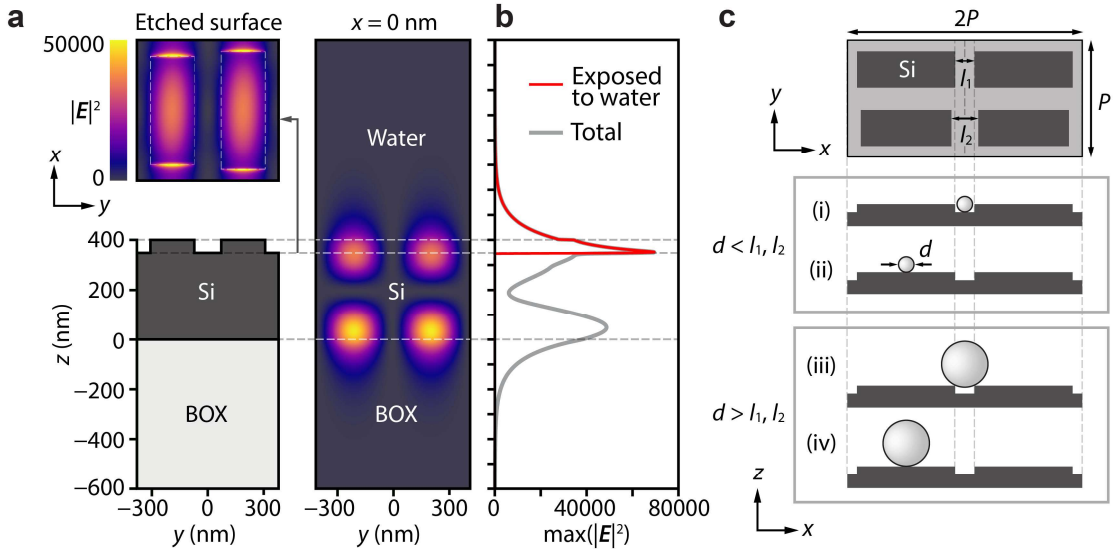

**Figure 3. Electric field enhancement and interaction mechanisms with nanoparticles.** (a) Cross-sectional schematic of the $yz$-plane (lower left) and simulated electric field intensity enhancement $|E^2|$ at $x = 0$ nm (right) with the field distribution in the $xy$-plane at a shallow-etched surface (upper left). (b) Maximum field intensity profile along the $z$-direction. (c) Schematic of two adjacent unit cells in the $xy$-plane and classification of nanoparticle binding configurations based on particle diameter $d$ and shallow-etched rod spacings $l_1$ and $l_2$.

We compared the peak wavelength shifts for different diameters of PS particles, as shown in Figure 4a. Resonance peak wavelengths were extracted in real-time from both raw data and simultaneous Fano fitting. When $d > l_1$ (~125 nm), the step height increased with increasing $d$; in contrast, the largest step height was observed at $d = 100$ nm, where $d < l_1$. This observation was confirmed by the average step heights shown in Figure 4b. As discussed above, this behavior can be attributed to the strong interaction between the localized electric field of the qBIC resonance and a single PS particle positioned between the shallow-etched nanorods. Although the Fano and raw peaks exhibited consistent trends, in some cases, only the raw peak showed a discrete step shift during the binding of small nanoparticles such as those with $d = 100$ nm. This can be attributed to changes in the Fano fitting parameters induced by local refractive index perturbations being too small to appear as a measurable resonance shift in real-time Fano fitting. Although the wavelength fluctuation 3σ of the raw peak (0.50 pm) was larger than that of the Fano peak (0.18 pm), local refractive index changes were readily identified as discrete steps in the raw spectra. Figure 4c presents a histogram of the step heights extracted from the raw peaks for $d = 100$ nm, yielding an average step height of 2.78 pm. The wavelength shift distribution followed a single-peak normal distribution; this is consistent with the typical statistics of single-particle binding events.[32] Further, we confirmed that the time interval Δ$t$ between successive binding events of PS particles follow a single-exponential decay, which is consistent with a Poissonian stochastic adsorption process[16] of individual binding events (Supporting Information S4).

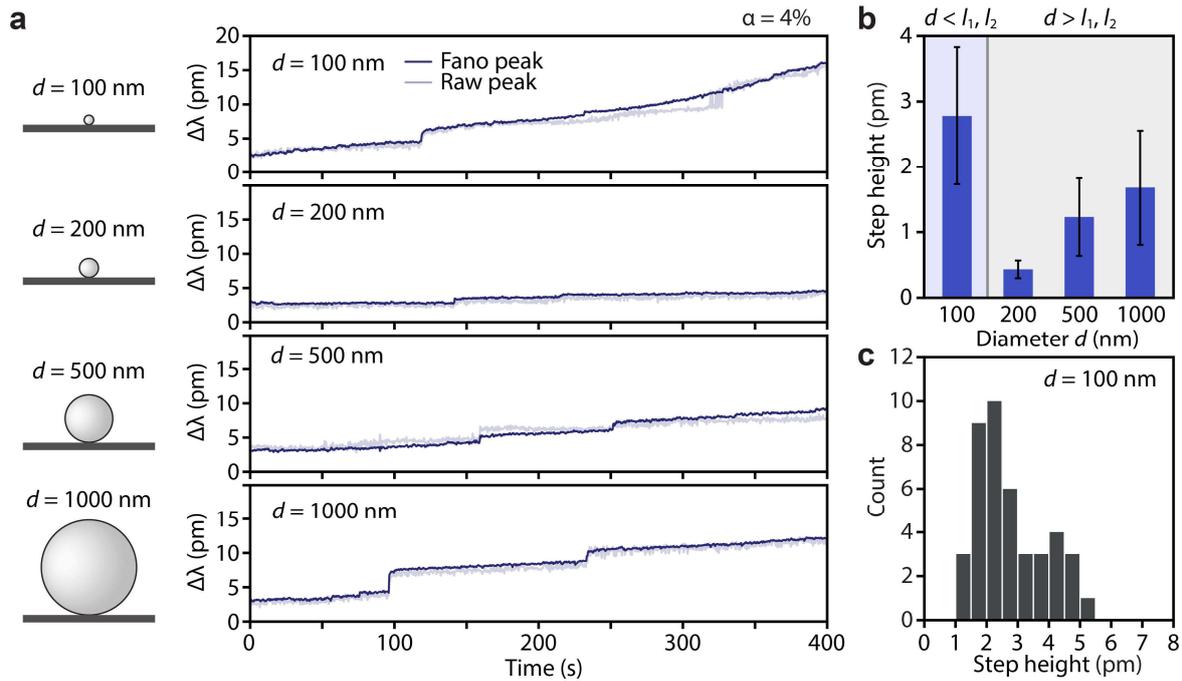

**Figure 4. Size dependence of PS particle binding.** (a) Real-time sensing results in the step-dominated region for different PS particle diameters *d*. Both Fano-fitted and raw spectral peaks are shown. (b) Average step height as a function of *d*. Error bars represent the standard deviation of the wavelength step heights. (c) Histogram of wavelength step heights for *d* = 100 nm.

**Binding-induced change in qBIC properties**

We investigated the effect of local refractive index perturbations on the resonance properties under qBIC conditions. As shown in Figure 5a, we analyzed how the peak wavelength, linewidth, and amplitude of the Fano resonance evolve upon single-particle binding. Figure 5b,c shows the real-time tracking results and corresponding histograms of the step heights, respectively, obtained after injecting a dispersion of unfunctionalized PS particles. Here, we selected a particle diameter of 500 nm to ensure sufficiently large signal changes. The solution pH was adjusted to 3.6 while maintaining a NaCl concentration of 100 mM to reduce electrostatic repulsion between the surface and the particles. Consequently, no transient trapping events were observed in peak wavelength, while multiple discrete steps were detected, all corresponding to redshifts, as shown in the histogram. The linewidth steps increased concurrently with the wavelength steps. The amplitude steps, on the other hand, exhibited both increase and decrease (the average step change Δ*A* was −0.013, indicating a slight decrease). Yang et al. recently demonstrated that variations in permittivity asymmetry can be encoded into the asymmetry factor of the system without modifying the geometric asymmetry.[33] Consequently, both the resonance position and amplitude become sensitive to changes in the surrounding refractive index. A similar mechanism applies in our case; the binding of single PS particles can modify the asymmetry factor of the qBIC resonance depending on the binding position of the particle. Specifically, when a particle binds at a location that increases (decreases) the effective asymmetry, the radiative coupling to the external continuum is enhanced (suppressed), leading to an increase (decrease) in the resonance amplitude. In these scenarios, the linewidth is expected to increase or decrease; however, only positive linewidth steps were observed in our experiments. This behavior may be related to weakened optical confinement or enhanced scattering induced by the binding of relatively high refractive index (*n* = 1.59) PS particles. Another notable observation is that, in some cases, no amplitude change was detected even when a wavelength step was observed. This behavior can be attributed to particle binding at locations where the change in effective asymmetry is insignificant. These results confirm that low-contrast BIC metasurfaces respond to local refractive index perturbations through simultaneous changes in three resonance parameters—wavelength, linewidth, and amplitude—all of which exhibit single-particle sensitivity with a high signal-to-noise ratio. Accumulating large datasets of such measurements and correlating them with physical parameters can help extract physical information, including particle size, position, and refractive index, utilizing artificial intelligence.

Beyond the sensing performance, silicon-based BIC metasurfaces offer an additional advantage in their compatibility with established complementary metal–oxide–semiconductor (CMOS) technologies,[34] enabling low-cost, precision-controlled mass production. Although this advantage is common to previously reported silicon photonic crystals,[35] spiral waveguides,[36] and silicon nitride microring resonators,[37] our silicon BIC metasurfaces enable single-nanoparticle detection using a simple free-space configuration in which normally incident laser light is transmitted through the device and measured directly. This approach eliminates the need for unstable fiber-based optical coupling systems[38] and fluorescence labeling,[21] thereby providing a more user-friendly sensing platform readily integrated with microfluidic systems.[39]

Although the present measurements rely on probabilistic single-event detection at relatively high particle concentrations, detection of trace-amounts of molecules or proteins will be more efficiently conducted through appropriate surface functionalization[40] such as aptamers[41] and by exploiting trapping potentials.[42] Further enhancement of the local electric field intensity can contribute to both signal enhancement and realization of trapping-assisted sensing based on self-induced back-action[9,17] by introducing nanogaps in paired shallow nanorods.[43,44]

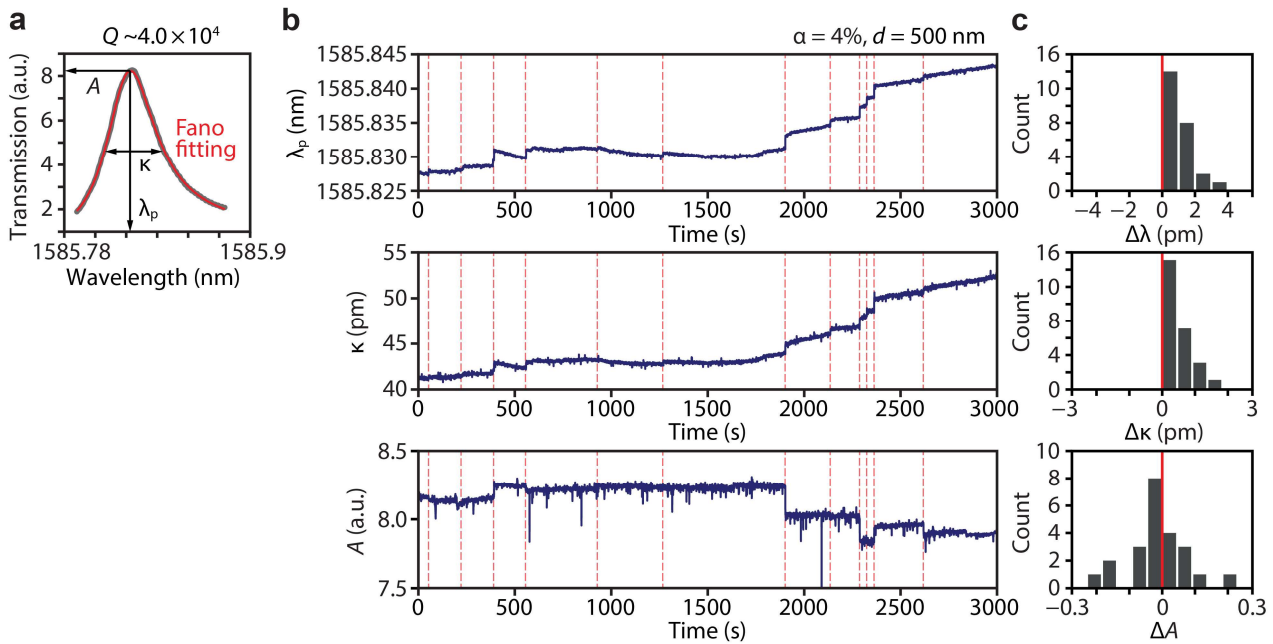

**Figure 5. Binding-induced changes in qBIC properties.** (a) Measured transmission spectrum and corresponding Fano fitting. (b) Real-time detection of single PS particles ($d$ = 500 nm) monitored via peak wavelength $\lambda_p$, linewidth $\kappa$, and peak amplitude $A$ (top to bottom). Red dashed lines indicate particle-binding events. (c) Histograms of step changes in peak wavelength $\Delta\lambda_p$, linewidth $\Delta\kappa$, and amplitude $\Delta A$ (top to bottom).

## ◼ Conclusions

We demonstrated, for the first time in BIC nanostructures, the detection of virus-sized single PS particles in water by leveraging the ultrahigh-$Q$ nature of a low-contrast geometry. This sensing capability arises from the strong interaction between the confined electric fields and nanoparticles located within the shallow-etched structures. We further revealed that the resonance properties of BIC metasurfaces—wavelength, linewidth, and amplitude—were modified by the binding of individual nanoparticles, all exhibiting single-particle sensitivity. This ultrahigh sensitivity can be understood as a consequence of binding-induced changes in the optical confinement and effective asymmetry of the qBIC system, representing a unique sensing feature enabled by BIC metasurfaces. Because of the collective resonant behavior of relatively large-area periodic arrays, BIC metasurfaces are inherently position-insensitive and do not require precise spatial alignment. In addition, simple normal-incidence excitation from free-space provides direct access to sharp qBIC resonances without complex optical coupling schemes. These advantages offer promising opportunities for simple, cost-effective, and highly sensitive detection of virions and other pathogenic nanoparticles. This platform could further enable real-time investigations of nanoscale binding dynamics and label-free biosensing at the ultimate single-entity limit.

## ■ Methods

**Fabrication.** The low-contrast BIC metasurfaces were fabricated on a SOI wafer consisting of a 400-nm-thick top silicon layer and a 2000-nm-thick buried oxide (BOX) layer. A 100-nm-thick resist AR-P 6200 was first spin-coated onto the wafer, followed by electron-beam lithography (ELS-BODEN, Elionix) to define metasurface patterns (100 μm x 100 μm). The patterns were transferred into the top silicon layer using a Bosch process (MUC-21 ASE-SRE, Sumitomo Precision Products) with $SF_6$ and $C_4H_8$ gases. The etching depth was controlled by adjusting the number of etching and passivation cycles to achieve the shallow-etched geometry. The resulting etching depth was measured using spectroscopic ellipsometry (M-2000, J.A. Woollam) and was approximately 52 nm.

**Characterization.** Unfunctionalized PS particles (Polysciences/Bangs Laboratories) were diluted in heavy water ($D_2O$, 151882, Sigma-Aldrich) containing 100 mM NaCl. The particle concentration was adjusted to 50 μg/mL for $d$ = 100 nm and 100 μg/mL for $d$ = 200, 500, and 1000 nm. The solutions were introduced into a custom-designed PDMS microfluidic channel (Micro TAS Engineering) via a silicone tube with an inner diameter of 1 mm. The output power of the tunable laser (TSL210V, santec) was set to 30–40 μW. The laser linewidth was intentionally broadened to 500 MHz to suppress interference noise during transmission spectrum acquisition. Signal steps of real-time measurement results were identified using the window-based change point detection algorithm in the *ruptures* Python library.[45] The wavelength fluctuation was defined as three times the standard deviation calculated from 20 consecutive measurement points.

**Simulation.** Electromagnetic field simulations were performed using the FDTD method (Ansys Lumerical). A single-unit-cell model was constructed with perfectly matched layer boundary conditions along the $z$-axis and Bloch boundary conditions along the $x$- and $y$- axes. The optical constants of silicon and the BOX ($SiO_2$) layer were taken from Palik.[46] Multiple electric dipole sources were randomly positioned at the mid-height of the top silicon layer ($z$ = 200 nm) to excite transverse electric (TE)-like modes. Subsequently, the radiative $Q$ factor $Q_r$ was calculated as $\omega_0 U(t)/P(t)$,[47] where $\omega_0$ is the resonance angular frequency, $U(t)$ is the stored electromagnetic energy, and $P(t)$ is the radiated power absorbed at the simulation boundaries. Spatial electric field profiles at the qBIC resonance were obtained using a three-dimensional field monitor under $x$-polarized plane-wave excitation.

## ■ Supporting Information

Characterization of the laser beam diameter; Transmission spectra of low-contrast BIC metasurfaces with different asymmetry parameters; Electric field profiles along the $z$-direction; Statistical analysis of PS particle binding process (PDF)

## ■ Acknowledgement

This work was financially supported by JSPS KAKENHI Grant Number (JP22K20496, JP24K17583), Iketani Science and Technology Foundation (Grant Number 0361252-A), and "Advanced Research Infrastructure for Materials and Nanotechnology in Japan (ARIM)" of the Ministry of Education, Culture, Sports, Science and Technology (MEXT). Proposal Number JPMXP1224NM5259, JPMXP1225NM5235. The authors thank Satoshi Ishii for his support with an experimental equipment.

## ■ Conflict of interest

The authors declare no competing interests.

**Supporting Information for**

**Single-nanoparticle detection using quasi-bound states in the continuum supported by silicon metasurfaces**

Keisuke Watanabe[1]*, Samuel Crowther[2], Masanobu Iwanaga[3], Frank Vollmer[2], and Tadaaki Nagao[1,4]

1.* International Center for Materials Nanoarchitectonics (MANA), National Institute for Materials Science (NIMS), 1-1 Namiki, Tsukuba, Ibaraki 305-0044, Japan.
2. Department of Physics and Astronomy, Living Systems Institute, University of Exeter, EX4 4QD, Exeter, UK.
3. Research Center for Electronic and Optical Materials, National Institute for Materials Science (NIMS), 1-1 Namiki, Tsukuba, Ibaraki 305-0044, Japan.
4. Department of Condensed Matter Physics, Graduate School of Science, Hokkaido University, Kita 10, Nishi 8, Kita-ku, Sapporo 060-0810, Japan.

*Email: watanabe.keisuke@nims.go.jp

**S1. Characterization of the laser beam diameter**

We employed the knife-edge method to characterize the beam profile and estimate the diameter of the focused laser beam on a silicon surface. A 40 µW laser beam was focused onto the silicon surface, and a beveled thin blade was positioned near the focal spot. As shown in Figure S1a, the blade was translated across the focused beam in 5 µm increments while the transmitted optical power was monitored using a photoreceiver. The spatial beam profile was derived by calculating the gradient of the measured power profile with respect to the blade position, as shown in Figure S1b. The resulting power gradient was well fitted by a Gaussian function. The beam diameter was determined as the full width at the $1/e^2$ maximum, yielding a beam diameter of 28.4 µm at the sample plane.

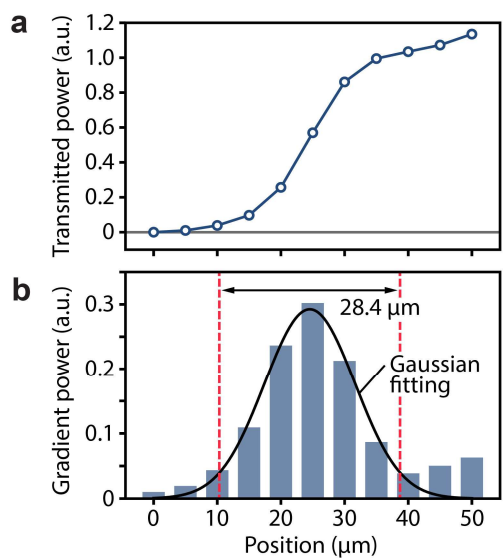

**Figure S1.** (a) Measured transmitted power as a function of knife-edge position. (b) Reconstructed beam profile together with the Gaussian fitting.

**S2. Transmission spectra of low-contrast BIC metasurfaces with different asymmetry parameters**

Figure S2a shows the simulated transmittance map, where the refractive index of the surrounding medium was set to 1.33. The resonance amplitude and linewidth increase with an increase in the asymmetry parameter α, while the resonance wavelengths remain constant. A metasurface array with different α values was integrated into a PDMS microfluidic channel (Figure S2b), and the transmission spectra were measured in $D_2O$ (Figure S2c). For α ≤ 3%, the laser output power was maintained at 50 µW. In this case, the resonance amplitude increased with increasing α. For α ≥ 4%, the laser output power was reduced to avoid detector saturation caused by the increased resonance amplitude, which confirmed that the amplitude continued to increase with α. Meanwhile, the resonance linewidth increased with α. These experimental observations were in good agreement with the simulated transmission spectra. Although slight fluctuations in the experimental resonance wavelength were observed for different α, the overall wavelength positions agreed well with the simulations.

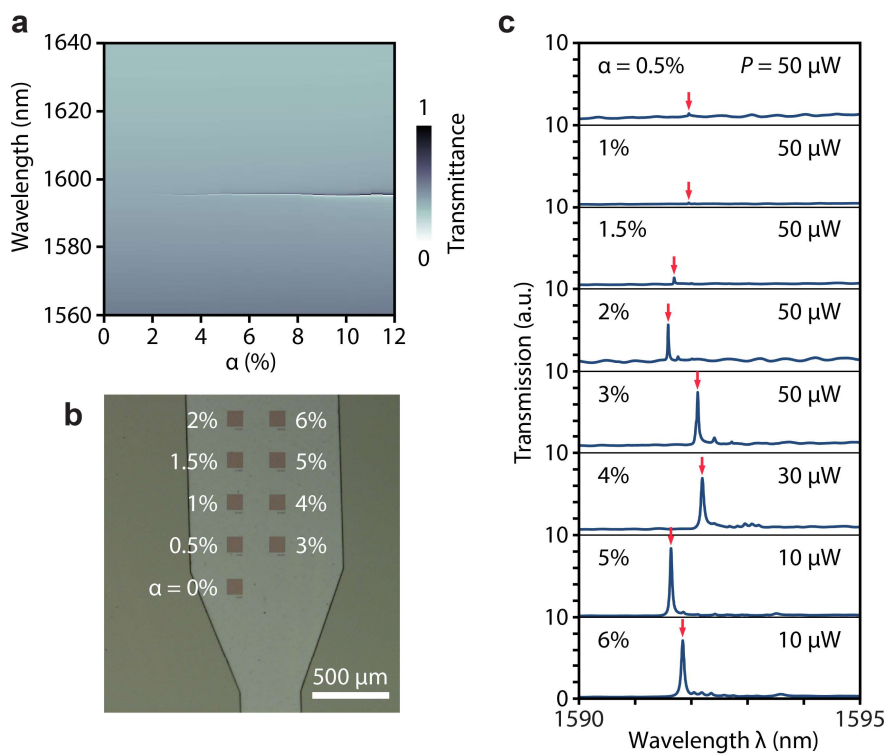

**Figure S2.** (a) Simulated transmittance map as a function of the asymmetry parameter α. (b) Optical microscope image of a fabricated device integrated with a PDMS microfluidic channel, showing a metasurface (100 µm x 100 µm) array with varied α. (c) Experimental transmission spectra measured for different *α*.

## S3. Electric field profiles along the *z*-direction

Figure S3 shows the simulated electric field enhancement in the *yz*-plane at *x* = 0 nm and *xy*-plane at different *z* positions. An *x*-polarized plane wave was normally incident on a metasurface, and the electric field distribution at the qBIC resonance wavelength was monitored. In the *yz*-plane, two antinodes appear along the *y*- and *z*- directions. The field profile of the *xy*-plane at the position of maximum electric field intensity along the *z*-direction ($z_{inmax}$) indicates a single antinode along the *x*-direction. These field characteristics confirm that the higher-order qBIC originates from a TE-like $TE_{122}$ mode. The confined optical mode partially penetrates the surrounding water because of the low-contrast geometry and large refractive index contrast between silicon and water. This results in strong electric field enhancement at the edges of the shallow-etched rods ($z_{surface}$). At the top silicon surface ($z_{top}$), the field enhancement at the rod edges becomes weaker and decays exponentially along the positive *z*-direction ($z_{500}$). These results confirm why the strongest light–molecule interactions occur near the shallow-etched regions.

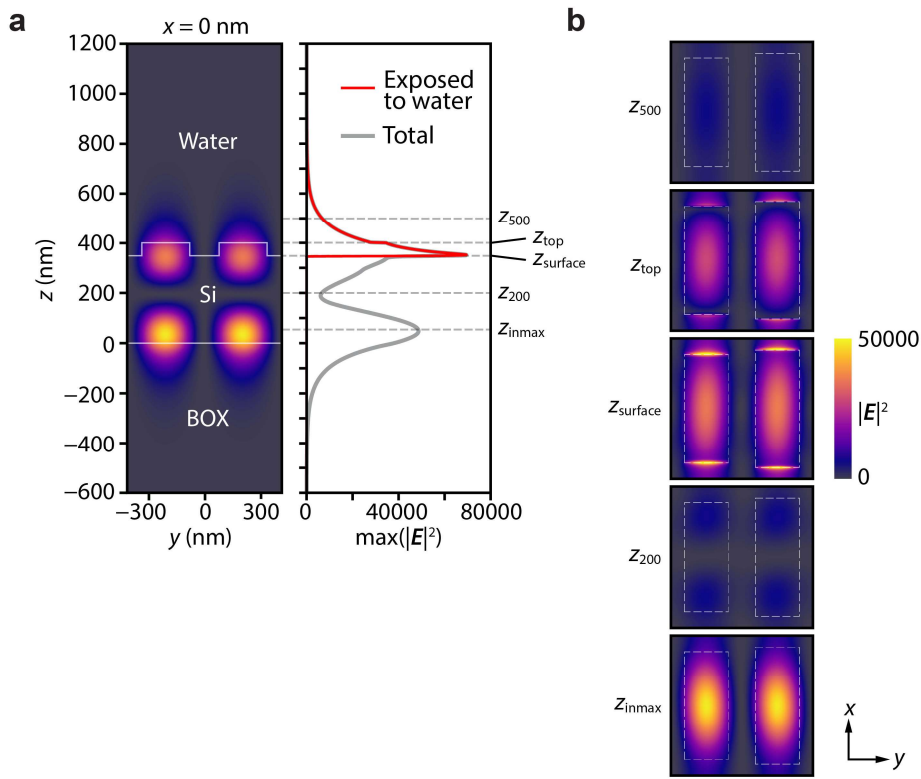

**Figure S3.** (a) Simulated electric field intensity $|E^2|$ at *x* = 0 nm in the *yz*-plane (left) and the corresponding maximum field intensity profile along the *z*-direction (right) for the metasurface with α = 4%. (b) Electric field distributions in the *xy*-plane at the different *z* positions ($z_{500}$, *z* = 500 nm; $z_{top}$, top silicon surface; $z_{surface}$, shallow-etched surface; $z_{200}$, *z* = 200 nm; $z_{inmax}$, position of maximum electric field intensity inside silicon).

## S4. Statistical analysis of PS particle binding process

Figure S4 shows the histogram of the time intervals Δ$t$ between successive binding events for PS particles with $d$ = 100 nm. The distribution follows a single-exponential decay, which is consistent with a Poissonian stochastic adsorption process. This statistical behavior indicates that individual binding events occur independently and probabilistically.

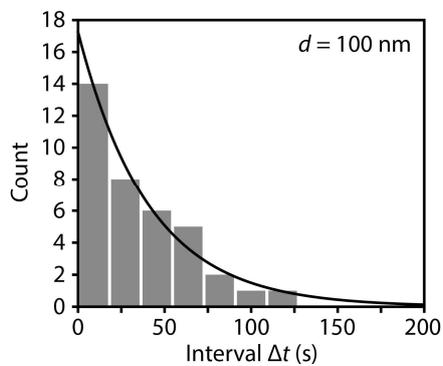

**Figure S4.** Histogram of the time intervals Δ$t$ between successive binding events of PS particles with $d$ = 100 nm. The black line represents an exponential fit.